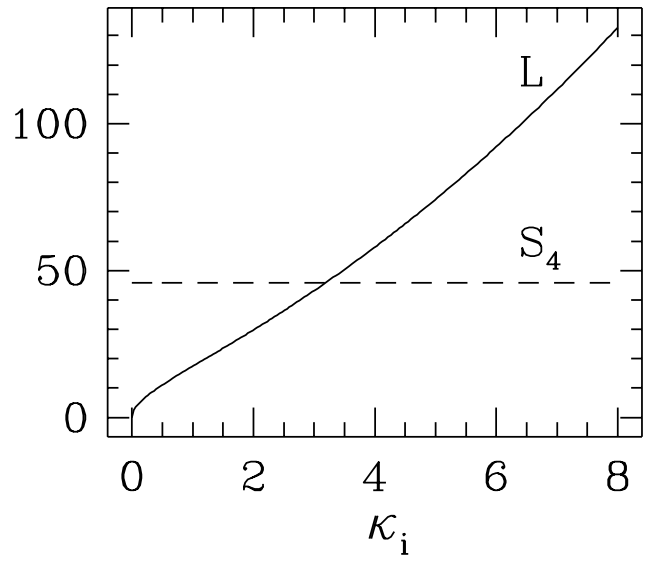

Figure 1.

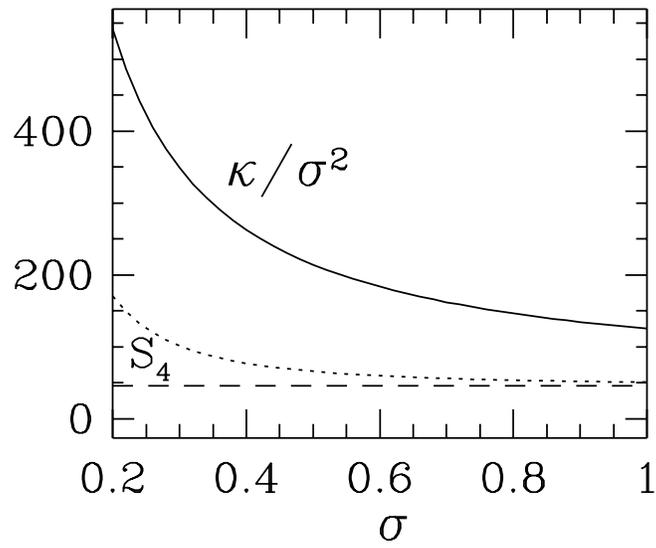

Figure 2.

# KURTOSIS IN LARGE-SCALE STRUCTURE
# AS A CONSTRAINT ON NON-GAUSSIAN INITIAL CONDITIONS

MICHAŁ J. CHODOROWSKI

*Institut d'Astrophysique de Paris, CNRS, 98 bis Bd. Arago*
*Paris, F-75014, France; chodor@iap.fr*

and

*Copernicus Astronomical Center, Bartycka 18*
*Warsaw, PL-00716, Poland; michal@camk.edu.pl*

and

FRANÇOIS R. BOUCHET

*Institut d'Astrophysique de Paris, CNRS, 98 bis Bd. Arago*
*Paris, F-75014, France; bouchet@iap.fr*

**ABSTRACT**

We calculate the kurtosis of a large-scale density field which has undergone weakly non-linear gravitational evolution from arbitrary non-Gaussian initial conditions. It is well known that the weakly evolved *skewness* is equal to its initial value plus the term induced by gravity, which scales with the rms density fluctuation in precisely the same way as for Gaussian initial conditions. As in the case of skewness, the evolved *kurtosis* is equal to its initial value plus the contribution induced by gravity. The scaling of this induced contribution, however, turns out to be qualitatively different for Gaussian versus non-Gaussian initial conditions. Therefore, measurements of the kurtosis can serve as a powerful discriminating test between the hypotheses of Gaussian and non-Gaussian nature of primordial density fluctuations.

## 1. Introduction

Redshift surveys of optically selected, and IRAS-selected galaxies have been recently used to estimate the distribution of galaxy counts on several scales (CfA and SSRS catalogs: Maurogordato & Lachièze-Rey 1987, Alimi et al. 1990, Gaztañaga 1992; 1.2Jy IRAS survey: Bouchet, Davis & Strauss 1992, Bouchet et al. 1993). The distribution turns out to be non-Gaussian, the departures from it being more pronounced for smaller scales. This fact does not necessarily imply that primordial density fluctuations were not Gaussian because nonlinear gravitational evolution does not preserve the shape of the fluctuations. Therefore, observed non-Gaussianity of the large-scale galaxy distribution may be either of primordial origin or due to subsequent nonlinear gravitational clustering, or the result of both.

Simple versions of inflation naturally produce Gaussian fluctuations from the quantum fluctuations of the inflaton field (Guth & Pi 1982; Hawking 1982; Starobinsky 1982; Bardeen, Steinhardt & Turner 1983). On the other hand, the density field generated by topological defects (Vilenkin 1985; Vachaspati 1986; Hill, Schramm & Fry 1989; Turok 1989; Bouchet & Bennett 1990; Bennett & Rhie 1990, Albrecht & Stebbins 1992) as well as in some versions of inflation involving multiple scalar fields (Allen, Grinstein & Wise 1987; Kofman & Pogosyan 1988; Salopek, Bond & Bardeen 1989) can be characterized as non-Gaussian. The comparison of predicted distributions of counts with existing observations, in either of the Gaussian or non-Gaussian cases, may serve not only as the confirmation of the gravitational instability theory, but possibly as a strong discriminating test between the two classes of models as well. It is, therefore, of great interest for cosmology to compute the predicted statistics of counts in both cases of initial conditions.

Numerous studies have been undertaken to calculate the gravity induced departures from an initial Gaussian distribution. Analytical efforts have been mostly successful in the weakly non-linear regime, where perturbation theory can be applied. The higher-order *reduced* moments (combinations of moments measuring departures from Gaussianity) of the evolved distribution have been calculated (Peebles 1980; Fry 1984; Goroff et al. 1986; Bouchet et al. 1992; Juszkiewicz, Bouchet & Colombi 1993; Bernardeau 1994a,b; Catelan & Moscardini 1994; Łokas et al. 1995) and recently the whole distribution function (Bernardeau 1992; Juszkiewicz et al. 1995; Bernardeau & Kofman 1995).

Given Gaussian initial conditions, the reduced moments, or cumulants, have been shown to obey the specific 'clustering hierarchy' (Fry 1984). In particular, skewness and kurtosis are expected to scale linearly and quadratically, respectively, with the square root of the variance of the density field, $\sigma$. (We use the standard statistical definition of skewness, $s$, and kurtosis, $\kappa$, namely $s = \langle \delta^3 \rangle / \sigma^3$, and $\kappa = \langle \delta^4 \rangle / \sigma^4 - 3$.) This is precisely what has been measured in redshift surveys. Furthermore, in the APM angular galaxy survey, containing as much as $\sim 1.3$ million galaxies, Gaztañaga (1994) has recently found to exist a 2D (angular) analog of the 3D hierarchy mentioned above. Bernardeau (1995), for Gaussian initial conditions, has indeed shown this angular hierarchy to be theoretically expected.

A non-Gaussian field has non-zero intrinsic, initial cumulants, even in the limit $\sigma \to 0$. Therefore, the observed linear scaling of skewness with $\sigma$ (Bouchet, Davis & Strauss 1992) has been used as a strong argument against non-Gaussian seeds of initial density perturbations (Silk & Juszkiewicz 1991; Coles & Frenk 1991). In particular, Silk & Juszkiewicz (1991) have argued that skewness is constant for the cosmic textures model, in conflict with the existing data. However, as pointed out by Coles *et al.* (1993, hereafter C93) and Luo & Schramm (1993, hereafter LS), a non-Gaussian initial perturbation must also undergo nonlinear gravitational evolution and the gravity-induced skewness may quickly dominate the initial term. Both groups have shown it to hold indeed true for some specific cases of non-Gaussian initial conditions: C93 using fully nonlinear N-body simulations, and LS in weakly nonlinear regime, applying perturbation theory. Fry & Scherrer (1994, hereafter FS) have calculated, in a simple and elegant way, the growth of skewness of the density field evolving weakly nonlinearly from *arbitrary* non-Gaussian initial conditions. FS have confirmed that in general the evolved skewness is equal to the initial, "increasingly unimportant" term plus the term which grows linearly with $\sigma$. They have concluded that "observations of a linear dependence of the skewness on the rms density fluctuation therefore do not necessarily rule out initially non-Gaussian models".

This is true. However, is it also the case for the cumulants of order higher than skewness? In other words, does the nonlinear gravitational dynamics wash away all information about the type (Gaussian or non-Gaussian) of initial conditions? The first, and so far, single attempt to answer this question has been done by LS who have also calculated the weakly evolved kurtosis. Unfortunately, the paper is unclear on this subject. In the abstract, "the importance of measuring the kurtosis is stressed since it will be preserved through the weakly nonlinear gravitational epoch". The formula (3.16) of LS for the evolved kurtosis, however, is a sensitive function of $\sigma$! This inspired us to come back to this problem in more detail.

In the present paper, we compute the kurtosis of a density field, which has undergone weakly nonlinear evolution from *arbitrary* non-Gaussian initial conditions. More specifically, we calculate the perturbation theory *lowest order* relevant contribution to the evolved kurtosis in its most general form. The FS method of the real-space analysis of skewness can be applied equally well to the case of kurtosis. Therefore, we do so; we also follow mostly their notation. The organization of the paper is as follows: in Section 2.1 we qualitatively discuss the effect of non-Gaussian initial conditions on the expected scaling of the cumulants of a density field with rms fluctuation. In Section 2.2 we derive the general formula for the evolved kurtosis. In Section 2.3 we calculate the effect of local biasing on this formula. In Section 3 we apply the formula to the scaled lognormal toy-model and give detailed predictions for the kurtosis – variance relationship in that case. Our conclusions are presented in Section 4.

## 2. Weakly Nonlinear Kurtosis

### 2.1. Qualitative Analysis

We are interested in the statistics of a large-scale density field, described by the density contrast, $\delta = [\rho(\mathbf{x},t) - \langle\rho\rangle]/\langle\rho\rangle$. For a Gaussian random process, all information is contained in the power spectrum, or its Fourier transform, the two-point correlation function, $\xi_2$. This, however, is no longer true for a non-Gaussian process. In terms of the p-point moments, the process contains non-vanishing p-point reduced functions, $\xi_p$, $p > 2$. By definition, $\langle\delta\rangle = 0$. The next few moments are

$$\langle\delta(\mathbf{x}_1)\delta(\mathbf{x}_2)\rangle = \xi_2(\mathbf{x}_1, \mathbf{x}_2),$$

$$\langle\delta(\mathbf{x}_1)\delta(\mathbf{x}_2)\delta(\mathbf{x}_3)\rangle = \xi_3(\mathbf{x}_1, \mathbf{x}_2, \mathbf{x}_3),$$

$$\langle\delta(\mathbf{x}_1)\delta(\mathbf{x}_2)\delta(\mathbf{x}_3)\delta(\mathbf{x}_4)\rangle = \xi_2(\mathbf{x}_1, \mathbf{x}_2)\xi_2(\mathbf{x}_3, \mathbf{x}_4) + \xi_2(\mathbf{x}_1, \mathbf{x}_3)\xi_2(\mathbf{x}_2, \mathbf{x}_4) +$$
$$\xi_2(\mathbf{x}_2, \mathbf{x}_3)\xi_2(\mathbf{x}_1, \mathbf{x}_4) + \xi_4(\mathbf{x}_1, \mathbf{x}_2, \mathbf{x}_3, \mathbf{x}_4),$$

$$\langle\delta(\mathbf{x}_1)\delta(\mathbf{x}_2)\delta(\mathbf{x}_3)\delta(\mathbf{x}_4)\delta(\mathbf{x}_5)\rangle = \xi_5'(\mathbf{x}_1, \mathbf{x}_2, \mathbf{x}_3, \mathbf{x}_4, \mathbf{x}_5) =$$
$$= \xi_2(\mathbf{x}_1, \mathbf{x}_2)\,\xi_3(\mathbf{x}_3, \mathbf{x}_4, \mathbf{x}_5) + \text{cycl. (10 terms)} +$$
$$\xi_5(\mathbf{x}_1, \mathbf{x}_2, \mathbf{x}_3, \mathbf{x}_4, \mathbf{x}_5)\,. \qquad (1)$$

In the above equation we have also defined an *unreduced* 5-point function, $\xi_5'$. The reasons why we have done so will become clear in the next Section. For a homogeneous and isotropic random process, the p-point averaged functions depend only on relative distances. For example, when $p = 2$, $\xi_2(\mathbf{x}_1, \mathbf{x}_2) = \xi_2(|\mathbf{x}_1 - \mathbf{x}_2|)$, and so on.

The cumulants of the *one-point* probability distribution function (PDF) are simply related to the p-point functions: the p-th cumulant, $\mathcal{K}_p = \xi_p(0)$, where $\xi_p(0) = \xi_p(0,\ldots,0)$. Throughout this paper we will be interested in *normalized* cumulants, or $\hat{\mathcal{K}}_p = \mathcal{K}_p/\sigma^p$, where

$$\sigma^2 = \langle \delta^2 \rangle \tag{2}$$

is the variance of the one-point PDF. The third and the fourth normalized cumulants are called in the literature skewness, $s$, and kurtosis, $\kappa$, respectively. For the fifth one, $q$, we coin here the name "pentosis". They are related to the central moments by the following equations:

$$s = \frac{\langle \delta^3 \rangle}{\sigma^3}, \tag{3}$$

$$\kappa = \frac{\langle \delta^4 \rangle}{\sigma^4} - 3, \tag{4}$$

and

$$q = \frac{\langle \delta^5 \rangle}{\sigma^5} - 10\, s\,. \tag{5}$$

For weakly nonlinear density perturbations ($\sigma < 1$) one can apply perturbation theory. Within the framework of perturbation theory one approximates the solution to gravitational dynamics equations for the density contrast as a series

$$\delta = \delta_{(1)} + \delta_{(2)} + \delta_{(3)} + \ldots, \tag{6}$$

where $\delta_{(p)}$ are found to be of order of $[\delta_{(1)}]^p$ (Fry 1984; Goroff et al. 1986). The first order (linear) term grows exclusively by an overall scale factor, $\delta_{(1)}(\mathbf{x}, t) = D(t)\, \delta(\mathbf{x}, t_i)$, where $\delta(\mathbf{x}, t_i)$ is the primordial fluctuation imprinted in the early Universe at an initial time $t_i$ and $D(t_i) \equiv 1$. Hence, in the linear regime ($\sigma \ll 1$), the probability distribution preserves its initial shape. In particular, its normalized cumulants, e.g. the skewness and kurtosis, of equation (4) and (5), remain unchanged. When higher order terms in equation (6) become non-negligible, they cause the cumulants to evolve. Since the lowest order contribution is given by

$$\langle \delta^p \rangle = \langle [\delta_{(1)} + \delta_{(2)} + \ldots]^p \rangle = \langle [\delta_{(1)}]^p \rangle + p\, \langle [\delta_{(1)}]^{(p-1)}\, \delta_{(2)} \rangle + \ldots, \tag{7}$$

it leads to

$$\langle \delta^2 \rangle = \langle [\delta_{(1)}]^2 \rangle + 2\, \langle \delta_{(1)}\, \delta_{(2)} \rangle + O([\delta_{(1)}]^4)\,, \tag{8}$$

$$\langle \delta^3 \rangle = \langle [\delta_{(1)}]^3 \rangle + 3\, \langle [\delta_{(1)}]^2\, \delta_{(2)} \rangle + \ldots, \tag{9}$$

$$\langle \delta^4 \rangle = \langle [\delta_{(1)}]^4 \rangle + 4\, \langle [\delta_{(1)}]^3\, \delta_{(2)} \rangle + O([\delta_{(1)}]^6)\,. \tag{10}$$

If the initial fluctuation field is Gaussian the second terms on the right hand side of equations (10) and (8) vanish. These terms are odd products of the initial variables and their average for a multivariate Gaussian is zero. This ensures that kurtosis, equation (4), for a density field evolved from Gaussian initial conditions scales quadratically with $\sigma$. However, for initially non-Gaussian fluctuations these terms cannot be expected to vanish in general. Therefore, in this case, we expect the lowest order contribution to the evolved kurtosis to scale *linearly* with the rms amplitude of the fluctuations.

On the other hand, in the above sense, the skewness is not the best statistics to discriminate between Gaussianity and non-Gaussianity: its lowest order correction (eq. [9]) is of even order, so it is already non-vanishing for Gaussian initial conditions. Indeed, it is clear, from equations (8) and (9), as found by FS, that the evolved skewness must be equal to its initial value plus a term proportional to $\sigma$; only the coefficient of proportionality is sensitive to the type of initial conditions. Unfortunately, the value of this coefficient cannot be used as a test for their nature. While perturbation theory describes the evolution of the mass distribution, observations probe the distribution of galaxies and there are many reasons for thinking that galaxies are *biased* tracers of mass. Fry & Gaztañaga (1994) have shown that if the galaxy density is a nonlinear but local function of the mass density then the galaxy density field possesses the same scaling hierarchy as the mass field (see also Juszkiewicz *et al.* 1995). Therefore, the agreement between the observed hierarchy of the cumulants of the galaxy field with the hierarchy predicted theoretically for the mass field is considered as a strong support for *both* assumptions of Gaussian initial conditions *and* local biasing. On the other hand, local biasing changes the hierarchical amplitudes. Any discrepancy between the derived from theory and the measured amplitudes can be, therefore, attributed entirely to biasing and it is impossible to deduce from them alone anything about initial conditions.

In sum, as already noted by LS, kurtosis seems to be a better test of the type of initial conditions. This led us to investigate the evolution of kurtosis in some detail.

*2.2. Quantitative Formula: Derivation*

The first term on the right hand side of equation (10) is, using the definition (4),

$$\langle [\delta_{(1)}]^4 \rangle = (3 + \kappa_i)\, \sigma_{(1)}^4. \tag{11}$$

Here, $\kappa_i$ denotes the initial ($\equiv$ first order) kurtosis. To calculate the next term, we need the second order contribution to the density contrast, $\delta_{(2)}$. Peebles (1980) derived

$$\delta_{(2)} = \frac{5}{7}\epsilon^2 - \epsilon_{,\alpha}\Delta_{,\alpha} + \frac{2}{7}\Delta_{,\alpha\beta}\Delta_{,\alpha\beta}, \tag{12}$$

where

$$\epsilon(\mathbf{x}) \equiv \delta_{(1)}(\mathbf{x}), \qquad \text{and} \qquad \Delta(\mathbf{x}) = \int \frac{d^3 x'}{4\pi} \frac{\epsilon(\mathbf{x}')}{|\mathbf{x} - \mathbf{x}'|}. \tag{13}$$

Hence,

$$4\left\langle [\delta_{(1)}]^3 \delta_{(2)} \right\rangle = \frac{20}{7}\langle \epsilon^5 \rangle - 4\langle \epsilon^3 \epsilon_{,\alpha}\Delta_{,\alpha}\rangle + \frac{8}{7}\langle \epsilon^3 \Delta_{,\alpha\beta}\Delta_{,\alpha\beta}\rangle. \tag{14}$$

The first term in the equation above is simply

$$\frac{20}{7}\langle \epsilon^5 \rangle = \frac{20}{7} q_i' \, \sigma_{(1)}^5, \tag{15}$$

where $q_i'$ is by construction the initial, normalized, fifth *central* moment. It is related to the pentosis (normalized fifth cumulant) of the initial field, $q_i$, by the equation

$$q_i' = q_i + 10\, s_i, \tag{16}$$

(see eq. [5]). The second and the third terms are

$$\begin{aligned}
-4\langle \epsilon^3 \epsilon_{,\alpha} \Delta_{,\alpha}\rangle &= -4 \int \frac{d^3 x'}{4\pi} \langle \epsilon^3(\mathbf{x}) \epsilon(\mathbf{x})_{,\alpha} \epsilon(\mathbf{x}')\rangle \frac{1}{|\mathbf{x} - \mathbf{x}'|_{,\alpha}} \\
&= -\int \frac{d^3 x'}{4\pi} \langle \epsilon^4(\mathbf{x}) \epsilon(\mathbf{x}')\rangle_{,\alpha} \frac{1}{|\mathbf{x} - \mathbf{x}'|_{,\alpha}},
\end{aligned} \tag{17}$$

and

$$\frac{8}{7}\langle \epsilon^3 \Delta_{,\alpha\beta}\Delta_{,\alpha\beta}\rangle = \frac{8}{7} \int \frac{d^3 x'\, d^3 x''}{4\pi\ 4\pi} \langle \epsilon^3(\mathbf{x}) \epsilon(\mathbf{x}') \epsilon(\mathbf{x}'')\rangle \frac{1}{|\mathbf{x} - \mathbf{x}'|_{,\alpha\beta}} \frac{1}{|\mathbf{x} - \mathbf{x}''|_{,\alpha\beta}}. \tag{18}$$

Now we apply the identity $\nabla_{,\alpha} f(\mathbf{x} - \mathbf{x}') = -\nabla'_{,\alpha} f(\mathbf{x} - \mathbf{x}')$ to equations (17) and (18). The integral in equation (17) can be then integrated by parts, giving

$$-4\langle \epsilon^3 \epsilon_{,\alpha}\Delta_{,\alpha}\rangle = -q_i' \, \sigma_{(1)}^5, \tag{19}$$

and equation (18) takes the form

$$\frac{8}{7}\langle\epsilon^3 \Delta_{,\alpha\beta}\Delta_{,\alpha\beta}\rangle = \frac{8}{7}\int \frac{d^3x'\,d^3x''}{4\pi\;4\pi}\,\xi_5{}'(0,0,0,\mathbf{x}',\mathbf{x}'')\,\frac{1}{|\mathbf{x}'|}_{,\alpha\beta}\,\frac{1}{|\mathbf{x}''|}_{,\alpha\beta}. \quad (20)$$

In the above, we have used equation (1) and the property of translational invariance. To simplify calculations we will carry them on with the unreduced 5-point function $\xi_5{}'$, and will substitute the partition of $\xi_5{}'$ at the very end.

Applying the identity

$$\nabla_\alpha \nabla_\beta \frac{1}{|\mathbf{x}|} = \frac{3\hat{x}_\alpha \hat{x}_\beta - \delta_{\alpha\beta}}{x^3} - \frac{4\pi}{3}\delta_{\alpha\beta}\,\delta_D(\mathbf{x}), \quad (21)$$

where $\delta_D(\mathbf{x})$ denotes the Dirac $\delta$ function, the integral in equation (20) can be rewritten as

$$\frac{1}{3}q_i{}'\,\sigma^5_{(1)} + \mathcal{C}[\xi_5{}']. \quad (22)$$

Here, the functional $\mathcal{C}$ is generally defined as

$$\mathcal{C}[f(\mathbf{x}',\mathbf{x}'')] = \int \frac{d^3x'\,d^3x''}{4\pi\;4\pi}\,\frac{3(3\cos^2\theta - 1)}{x'^3 x''^3}\,f(\mathbf{x}',\mathbf{x}''), \qquad \cos\theta = \frac{\mathbf{x}'\cdot\mathbf{x}''}{x'x''}. \quad (23)$$

Let us introduce the normalized functional, $\hat{\mathcal{C}}$, defined by

$$\hat{\mathcal{C}}[f] = \mathcal{C}[f]/f(0,0). \quad (24)$$

Equation (20) can then be finally expressed as

$$\frac{8}{7}\langle\epsilon^3 \Delta_{,\alpha\beta}\Delta_{,\alpha\beta}\rangle = \frac{8}{7}\left(\frac{1}{3} + \hat{\mathcal{C}}[\xi_5{}']\right)q_i{}'\,\sigma^5_{(1)}. \quad (25)$$

Combining the equations (10), (11), (14), (15), (19), and (25), we obtain

$$\langle\delta^4\rangle = (3 + \kappa_i)\,\sigma^4_{(1)} + a'\,q_i{}'\,\sigma^5_{(1)}, \quad (26)$$

where

$$a' = \frac{47}{21} + \frac{8}{7}\hat{\mathcal{C}}[\xi_5{}']. \quad (27)$$

Similar calculations for $\sigma^2 = \langle\delta^2\rangle$ yield (see eq. [18] of FS)

$$\sigma^2 = \sigma^2_{(1)} + \frac{1}{2}b\,s_i\,\sigma^3_{(1)}, \quad (28)$$

where $s_i$ is the initial skewness and

$$b = \frac{26}{21} + \frac{8}{7}\hat{\mathcal{C}}[\xi_3]. \tag{29}$$

From equations (4), (26) and (28), we obtain the final expression for the evolved kurtosis:

$$\kappa = \kappa_i + L\,\sigma + O(\sigma^2)\,. \tag{30}$$

Here,

$$L = a'\,q_i{}' - b\,(3 + \kappa_i)\,s_i\,, \tag{31}$$

with $a'$ and $b$ given by equations (27) and (29).

Let us now recall that $\xi_5{}' = \xi_5 + \xi_2(1,2)\,\xi_3(3,4,5) + \mathrm{cycl.}$ (see eq. [1]). Thus, we have

$$\begin{aligned}\xi_5{}'(0,0,0,\mathbf{x}',\mathbf{x}'') =& \xi_5(0,0,0,\mathbf{x}',\mathbf{x}'') + \\ & 3\,\xi_2(\mathbf{x}')\,\xi_3(\mathbf{x}'') + 3\,\xi_2(\mathbf{x}'')\,\xi_3(\mathbf{x}') + \\ & 3\,\xi_2(0)\,\xi_3(0,\mathbf{x}',\mathbf{x}'') + \xi_2(\mathbf{x}' - \mathbf{x}'')\,\xi_3(0)\,, \end{aligned} \tag{32}$$

with $\xi_p(\mathbf{x}) = \xi_p(|\mathbf{x}|) \equiv \xi_p(0,\ldots,\mathbf{x})$. Let us calculate $\mathcal{C}[\xi_5{}']$ using the above partition of $\xi_5{}'$. The first term is simply $\hat{\mathcal{C}}[\xi_5]$. The next two terms vanish by symmetry. The fourth one is $3\,\mathcal{C}[\xi_3]\,\sigma_{(1)}^2$, and our calculation of the fifth one yields $\frac{2}{3}s_i\,\sigma_{(1)}^5$.

By using equation (16) we can cast the formula for $L$, equation (31), in the following form:

$$L = a\,q_i + \left[\frac{136}{7} - b\,\kappa_i\right]s_i\,, \tag{33}$$

where

$$a = \frac{47}{21} + \frac{8}{7}\,\hat{\mathcal{C}}[\xi_5]\,, \qquad \text{and} \qquad b = \frac{26}{21} + \frac{8}{7}\,\hat{\mathcal{C}}[\xi_3]\,. \tag{34}$$

The formula (30) for the evolved kurtosis, with $L$ given by equations (33)-(34), constitutes the main result of this Section. It shows that the weakly evolved kurtosis is not equal, as one might naïvely expect, to the initial kurtosis plus just the term proportional to the *squared* variance. Instead, as indicated by qualitative analysis in the previous Subsection, the term *linear* in variance is already present.

2.3. *Bias and Kurtosis*

Perturbation theory predicts the statistics of the *mass* fluctuations, while observations probe the *galaxy* distribution. There are both theoretical arguments and observational evidence for the idea that galaxies are *biased* tracers of the mass distribution. Without discussing the general issue of biasing in detail, in the present Subsection we calculate the weakly evolved kurtosis of galaxy distribution, $\kappa_g$, in the local bias model (Fry & Gaztañaga 1994; see also Juszkiewicz *et al.* 1995).

In this model, one assumes the galaxy density $\delta_g$ to be a nonlinear, but local function of the mass density,

$$\delta_g = f(\delta). \tag{35}$$

The Taylor expansion for $\delta_g$ up to second order is

$$\delta_g = b\,\delta + \tfrac{1}{2}b_2\,\delta^2 - \tfrac{1}{2}b_2\,\sigma^2, \tag{36}$$

where

$$b = f'(0), \qquad b_2 = f''(0), \qquad \sigma^2 = \langle \delta^2 \rangle. \tag{37}$$

The last term on the right-hand side of equation (36) ensures that $\langle \delta_g \rangle = 0$. The calculation of the second and the fourth central moment of $\delta_g$, up to the lowest order nonlinear correction, yields

$$\sigma_g^2 = \langle \delta_g^2 \rangle = b^2 \sigma^2 \left[ 1 + \frac{b_2}{b} s_i\, \sigma + O(\sigma^2) \right], \tag{38}$$

and

$$\langle \delta_g^4 \rangle = b^4 \sigma^4 \left[ 3 + \kappa_i + \left( L + 2\frac{b_2}{b}(q_i' - s_i) \right) \sigma + O(\sigma^2) \right]. \tag{39}$$

In deriving the above, we have made the substitutions $\langle \delta^3 \rangle = s_i \sigma^3 + O(\sigma^4)$, $\langle \delta^4 \rangle = (3+\kappa_i)\,\sigma^4 + L\,\sigma^5 + O(\sigma^6)$ (eq. [30]), and $\langle \delta^5 \rangle = q_i'\sigma^5 + O(\sigma^6)$. The kurtosis in galaxy distribution can be therefore calculated, using definition (4) and equation (16), as

$$\kappa_g = \kappa_i + L_g\, \sigma_g + O(\sigma_g^2), \tag{40}$$

where

$$L_g = \frac{L}{b} + 2\,\frac{b_2}{b^2}\Big[q_i + (6 - \kappa_i)\,s_i\Big], \tag{41}$$

and $L$ is given by equations (33)-(34). Thus the local biasing preserves the scaling of the weakly evolved kurtosis. The coefficient $L_g$ is a function of the normalized cumulants of the initial mass fluctuations field and of the coefficients $b$ and $b_2$. In principle, there are combinations of $b$ and $b_2$ for which $L_g$ vanishes. It is, however, quite unlikely, because it requires a coincidental cancellation of the terms of different origin, namely the $L/b$ term, arising from non-Gaussian initial conditions, by the $b_2$ term, arising from nonlinear biasing.

## 3. Applications: Scaled Lognormal Model

The simplest non-Gaussian models are purely local: a one-point PDF is non-Gaussian, but p-point correlations vanish, except at zero lag. Although physically not realistic, these models are very useful – due to their simplicity – to demonstrate the difference between the evolution of the kurtosis of initially Gaussian, and non-Gaussian density fields. This class of models has been investigated analytically by Fry & Scherrer (1994) and numerically by Messina *et al.* (1990) and Weinberg & Cole (1992). In the present paper we will restrict ourselves to the investigation of such models.

In these models, $\hat{\mathcal{C}}[\xi_3] = \hat{\mathcal{C}}[\xi_5{}'] = 0$. Therefore, to calculate the coefficient $L$ in the general formula for the kurtosis (eq. 30), it is more convenient to use the form given by equation (31). The coefficients $a'$ and $b$ (see eq. [27] and [29]) are thus $a' = 47/21$ and $b = 26/21$. By using additionally (16), we obtain

$$L = \frac{1}{21} \left[ 47\, q_i + (392 - 26\, \kappa_i)\, s_i \right]. \tag{42}$$

As a model of one-point distribution function, we now use the (scaled) lognormal distribution. The PDF of the lognormal distribution, $\Lambda(\Sigma, \nu)$, has the form:

$$P(y)\, dy = \frac{1}{\Sigma \sqrt{2\pi}} \exp\left[-\frac{(\log y - \nu)^2}{2 \Sigma^2}\right] \frac{dy}{y}. \tag{43}$$

For the value of the parameter $\nu$ we choose

$$\nu = -\Sigma^2/2. \tag{44}$$

The moments of such a distribution about the origin are

$$\mu'_n = \tau^{[n(n-1)/2]}, \qquad \tau = \exp\left[\Sigma^2\right]. \tag{45}$$

The initial density contrast, $\epsilon$, is related to the random variable $y$ by the equality

$$\epsilon = c\,(y - 1). \tag{46}$$

The condition (44) ensures that $\langle \epsilon \rangle = 0$. Indeed, $\langle \epsilon \rangle = c \langle y - 1 \rangle = c\,(\mu'_1 - 1)$, which, from equation (45), is equal to zero. The coefficient $c$ is uniquely determined by the condition

$$\langle \epsilon^2 \rangle = \sigma_i^2, \tag{47}$$

which gives $c = (\tau - 1)^{-1/2} \sigma_i$. Higher-order moments of $\epsilon$ are given by

$$\langle \epsilon^n \rangle = c^n \langle (y-1)^n \rangle, \tag{48}$$

and can be easily expressed in terms of the moments of the variable $y$, equation (45). The resulting initial skewness, kurtosis, and fifth normalized cumulant are respectively:

$$\begin{aligned} s_i &= (\tau - 1)^{1/2} (\tau + 2), \\ \kappa_i &= (\tau - 1)(\tau^3 + 3\tau^2 + 6\tau + 6), \\ q_i &= (\tau - 1)^{3/2} (\tau^6 + 4\tau^5 + 10\tau^4 + 20\tau^3 + 30\tau^2 + 36\tau + 24). \end{aligned} \tag{49}$$

Here, $\tau > 1$ is a free parameter of the model (see the definition of $\tau$ in eq. [45]). In this manner, we have constructed an initial field with a given initial rms fluctuation $\sigma_i \ll 1$, which can be arbitrarily far from Gaussian. This is different from the context in which the lognormal model has been introduced in cosmology: Coles & Jones (1991) have proposed this model as an approximation of the PDF evolution from *Gaussian initial conditions*. Indeed, for $\Sigma \to 0$, the distribution (43) tends to Gaussian. (Note that consequently the cumulants, equation [49], then tend to zero.) Therefore, to avoid possible misunderstanding, we refer to this model by the name Scaled Lognormal Model.

The higher-order cumulants of our distribution are determined uniquely by the value of $\tau$, so anyone of them can serve as a parametrization of the field. In the following, we parametrize the initial field by the value of its kurtosis, $\kappa_i$. In Figure 1, we plot the coefficient $L$ of equation (42), as a function of $\kappa_i$. (For simplicity, we investigate the non-bias case, so $L_g = L$ and $\kappa_g = \kappa$.) If the density field evolved from Gaussian initial conditions, then $L = 0$ and the first nonvanishing term in the formula for the evolved kurtosis, equation (30), would be that proportional to $\sigma^2$. In the present paper we have not attempted to compute the proportionality constant of this term in the case of non-Gaussian initial field. Still, for reference, in Figure 1 we plot the value of this coefficient for Gaussian initial conditions, $S_4 = 60712/1323 = 45.9$.

In Figure 2, we plot $\kappa/\sigma^2$ as a function of $\sigma$. The dashed line represents the Gaussian initial conditions prediction, i.e. $\kappa/\sigma^2 = S_4$. The solid line is drawn according to the formula (30) for a non-Gaussian field, for $\kappa_i = 5$. Here, $S_4$ has been accepted as the proportionality constant of the quadratic term. Finally, the dotted line also corresponds to equation (30), but with the '$L\sigma$' term set to zero. From the plot one can see that it could be difficult to distinguish observationally between the dotted and the dashed curve, due to big errors in the small $\sigma$ part of the diagram (Bouchet et al. 1993; Gaztañaga 1994). On the contrary, the difference between the solid and the dashed curve is obvious. This illustrates the importance of the linear-in-$\sigma$ term in the formula for the evolved kurtosis from a non-Gaussian initial field. This term ensures that the kurtosis is very sensitive to the type of initial conditions. In particular, our toy model with $\kappa_i \geq 5$ seems to be in clear conflict with observational data.

It should be emphasized that our results hold firm in spite of the fact that we have not calculated the coefficient of the $O(\sigma^2)$ contribution to the evolved kurtosis. The departure of the value of this coefficient from the value we used, $S_4$, can only shift the solid line in Figure 1 up or down, and in any case the solid and the dashed lines will remain distinctly different. (Even if we calculated the quadratic, $O(\sigma^2)$ term, the final controversy would remain about biasing, which can change the amplitude of this term for *galaxy* counts. Again, unless the biasing accidentally happens to cancel the linear, $L$ term (see eq . [41]), the Gaussian, and the non-Gaussian initial conditions predictions will remain qualitatively different.)

The lognormal distribution for the initial fluctuation field has been applied by Moscardini et al. (1993) to study, by means of N-body simulations, the evolution of the two-point correlation function from non-Gaussian initial conditions. In the present paper we are not trying to impose any constraints on the models investigated in that work, for a number of reasons. First, the lognormal distribution has been used by Moscardini et al. (1993) to model the gravitational potential fluctuation field, not the density one. Second, the initial field also had there lognormal correlations, absent in the toy model we investigated. And last but not least, we have performed our calculations for the case of an *unsmoothed* final field. However, in order to compare quantitatively the perturbation theory predictions with both N-body and observations, the effects of smoothing must be taken into account.

The calculation of the effect of smoothing has proven to be a difficult task already for Gaussian fluctuations (Juszkiewicz, Bouchet & Colombi 1993; Padmanabhan & Subramanian 1993; Bernardeau 1994a,b; Catelan & Moscardini 1994; Łokas et al. 1995). For a non-Gaussian initial field this is even more difficult, and we address this problem elsewhere (Catelan et al. 1995). Furthermore, the effect of the final smoothing of an initially non-Gaussian field is model dependent, since spatial averages will then depend not only on the initial power spectrum, but on higher order reduced p-point correlation functions as well. Nevertheless, we think that smoothing cannot change the qualitative picture emerging from our calculations: being a linear transformation of the final field, this cannot add or remove terms in the formula for the evolved kurtosis, equations (30)-(33).

## 4. Conclusions

In this paper, we have calculated the kurtosis of a density field which had undergone weakly nonlinear evolution from *arbitrary* non-Gaussian initial conditions. We have computed the perturbation theory *lowest order* relevant contribution to the evolved kurtosis.

This contribution is in general *linear* in the rms amplitude of the density fluctuations, but it is absent for Gaussian initial conditions. Therefore, it induces different scaling of kurtosis with $\sigma$ from that for Gaussian fluctuations. We have shown this difference to be very distinct for the scaled lognormal toy-model.

Galaxies are most likely biased tracers of mass. The local biasing preserves the structure of the Gaussian hierarchy of cumulants. The accordance with observed hierarchy in galaxy counts is therefore regarded as a strong support for both assumptions of Gaussian initial conditions *and* local biasing. On the other hand, the coefficients of the scaling (the amplitudes) of cumulants are substantially changed. Therefore, they cannot be used as a test of the nature of initial conditions. This fact additionally strengthens the importance of the different scaling of kurtosis.

The subject deserves further study, the most important step being to take into account the effect of smoothing of the final field. However, as argued in Section 3, it is unlikely that smoothing can change qualitatively the picture outlined above. Still, even if it happened to be so, it would not be *dynamics* which wipes out the information about the type of initial conditions! In the present paper we have shown that weakly nonlinear gravitational clustering preserves the information about initial conditions, and that kurtosis is a good indicator of this fact.

## Acknowledgments

We are grateful to Francis Bernardeau and Paolo Catelan for helpful comments. M. Chodorowski acknowledges partial support by a fellowship from the French Ministère de l'Enseignement Supérieur et de la Recherche and by the Polish Government Committee for Scientific Research (KBN) grant number 2P30401607.

# Figure Captions

**Figure 1.** The coefficient $L$, i.e. the coefficient of the linear-in-$\sigma$ term in the formula for the evolved kurtosis (cf. equation 30) as a function of the initial kurtosis, $\kappa_i$, for the Scaled Lognormal Model. For reference, the value of the coefficient of the term quadratic in $\sigma$ for Gaussian initial conditions (first nonvanishing one in this case), $S_4$, is also marked.

**Figure 2.** The evolved kurtosis over evolved variance, $\kappa/\sigma^2$, as a function of $\sigma$ for the Scaled Lognormal Model. The dashed line represents the Gaussian initial conditions prediction, i.e. $\kappa/\sigma^2 = S_4$. The solid line is drawn according to the formula for a non-Gaussian field, equation (30) with $L$ given by equation (42), for $\kappa_i = 5$. Here, $S_4$ has been taken as the proportionality constant of the quadratic term. Finally, dotted line is also drawn from equation (30), but with the '$L\sigma$' term set to zero.